\documentclass[12pt]{article}
\usepackage{epsfig}
\def\ltsim{\lower3pt\hbox{$\, \buildrel < \over \sim \, $}}
\def\gtsim{\lower3pt\hbox{$\, \buildrel > \over \sim \, $}}
\def\be{\begin{equation}}
\def\ee{\end{equation}}
\def\ba{\begin{eqnarray}}
\def\ea{\end{eqnarray}}
\def\ga{\mathrel{\raise.3ex\hbox{$>$\kern-.75em\lower1ex\hbox{$\sim$}}}}
\def\la{\mathrel{\raise.3ex\hbox{$<$\kern-.75em\lower1ex\hbox{$\sim$}}}}

\begin{document}

\baselineskip=16pt
\begin{titlepage}

\begin{center}

\vspace{0.5cm}

\Large{\bf Gravity In A Model Of Multi Branes Alternated Between Minkowski
And AdS Spacetime 
In The Fifth Dimension  }

\vspace{10mm}
Yun-Song Piao$^{a,c}$, Xinmin Zhang$^{a}$ and Yuan-Zhong Zhang$^{b,c}$ \\

\vspace{6mm}

{\footnotesize{\it
 $^a$Institute of High Energy Physics, Chinese
     Academy of Sciences, P.O. Box 918(4), Beijing 100039, China\\
 $^b$CCAST (World Lab.), P.O. Box 8730, Beijing 100080}\\
 $^c$Institute of Theoretical Physics, Chinese Academy of Sciences,
      P.O. Box 2735, Beijing 100080, China \footnote{Mailing address in
      China, Email address: yspiao@itp.ac.cn}\\}

\vspace*{5mm}
\normalsize
\smallskip
\medskip

\smallskip
\end{center}
\vskip0.6in

\centerline{\large\bf Abstract}
 {We construct in this paper a multi-brane model with
Minkowski and AdS spacetime alternated in the fifth dimension
and study the gravity on the brane of our universe. We will show that in
this model there are
both the graviton-like resonant state which generates the
``quasi-localized gravity" and the resonant states of the massive
KK mode.  This model gives rise to gravity which deviates from the Newton
gravitational potential at small and large scale which, however can be
shown to be consistent with the
observations in
the absence of fine-tuning of the model parameters.
 }

\vspace*{2mm}

\end{titlepage}

As motivated by the string theory \cite{HW} \cite{WL}, there has been
a lot of studies in the recent years on brane models.
These models could provide a solution to
problem of the hierarchy between the
Plank scale and the electroweak scale \cite{ADD} \cite{RS}.
Furthermore various generalizations of models in \cite{ADD} \cite{RS},
such as the multi-brane
models \cite{GRS} \cite{KR} and Crystal Universe model \cite{K}
\cite{YS}, have also been proposed and studied in the literature.
Phenomenologically these
models provide a way of placing the visible sector on a positive tension
brane and give rise to many predictions
\cite{ADDK}on
neutrino mixing and Dark Matter. Usually the branes are supposed to live
in the AdS spacetime.
Recently the authors of Refs.\cite{GRS} and
\cite{CEH}
have
proposed multi-brane models with ``quasi-localized gravity" where
five dimensional (5D) Minkowski region is embedded in AdS region and
the
gravity is mediated not
by a normalized graviton state but by a resonant state. In the model of
Ref. \cite{LMW}, a finite Minkowski
region is surrounded by a AdS region and the
resonant modes modify gravity at
small scale.

In this paper we study a brane setup which is a generalization of the
multi-brane model considered in 
\cite{YS} with Minkowski and
AdS spacetime alternated in the fifth dimension, then calculate the
gravitational potential between
two unit masses on positive tension brane.
In our model, there is not only a zero mode resonant state which
generates the "quasi-localized gravity", but also exist
many massive resonant states 
of the massive KK mode, which modify Newtonian
gravitational potential at different length scales.

Our 5D model is shown in Fig. 1 which consists of
four parallel $3$-branes \footnote{
This model can easily be generalized to include more than four $3$-branes.}
with Minkowski and AdS
spacetimes alternated and a $z_2$
symmetry in the fifth dimension $y$.
Due to the $z_2$ symmetry we
consider $y\geq0$. Two $3$-branes are located at $y=y_1$ and
$y=y_2$ and the action
of this brane setup is

\be S_{\rm bulk} = \int d^4x\int \sqrt{-{g}}
\left({1\over 2\kappa^2} R-\tilde{\Lambda}\right)dy~~~%
\mathrm{with}~~~\tilde{\Lambda}=\left\lbrace \matrix{0 &{\rm for}&
0\leq y_1&{\rm and}&y\geq y_2 \cr \Lambda &{\rm for}& y_1\leq
y\leq y_2} \right. \ee \be S_{\rm brane} = -\int d^4x
\left(\sqrt{-g_{1}}V_1+\sqrt{-g_{2}}V_2\right),
\ee
where $\kappa$ sets the 5D fundamental scale, $g_{1}$ and $g_{2}$
are the induced metrics on the branes located at $y=y_1$ and $y=y_2$
respectively, and the corresponding brane tensions denoted by $V_1$ and
$V_2$ are $\Lambda/2k$ and $-\Lambda/2k$ . The 5D
metric ansatz which respects 4D Poincare invariance is given by
\be
ds^2 = e^{-2\sigma(y)}\eta_{\mu\nu}dx^{\mu}dx^{\nu} + dy^2,
\ee
where the warp function $\sigma(y)$ is essentially a conformal
factor that rescales the 4D component of the metric:
$\sigma(y)=ky$ for $y_1\leq y\leq y_2$ and
$\sigma(y)$ equal to $k y_1$ and $k y_2$ for $ 0 \leq y \leq y_1$ and
$y_2 \leq y$
respectively, where
$k=\sqrt{\frac{-\kappa^2\Lambda}{6}}$ is effectively the bulk curvature
in the region between the two $3$-branes located at $y=y_1$ and $y=y_2$.

Let's consider the linearized gravitational perturbations about
the background metric in Eq.(3):
\be
ds^2= (e^{-2\sigma(y)}\eta_{\mu
\nu}+ h_{\mu \nu})dx^{\mu}dx^{\nu} +
 dy^2\ .
\ee
Following Ref. \cite{GW1}, we work in the transverse-traceless
gauge, $\partial^{\mu}h_{\mu \nu} =h^{\mu}_{\mu}=0$,  and do not
consider the metric fluctuations $h_{55}$ and $h_{5\mu}$. It is
useful to define a conformal coordinate
\be
z=\left \lbrace
\matrix{{e^{ky_1}-1\over k}&{\rm for}& 0 \leq y\leq y_{1}\cr
{e^{ky}-1\over k}&{\rm for}& y_{1}\leq y\leq y_{2}\cr
{e^{ky_2}-1\over k}&{\rm for}& y\geq y_{2}. }
\right.
\ee
We use the ansatz $h_{\mu \nu}=e^{ip \cdot x}
\exp{\left(-{\sigma(z)\over 2} \right)\Psi_m(z) \epsilon_{\mu
\nu}}$ with $\epsilon_{\mu \nu}$ being a constant
polarization tensor and $m^2=-p^2$ is an effective 4D mass.
The function ${\Psi}_m(z)$ obeys the following
Schr\"{o}dinger-like equation which is derived from the linearized Einstein
equation:

\be
\left[ -\frac{1}{2} \partial_{z}^2 + V(z) \right] \Psi_m(z) =
\frac{1}{2}m^2 \Psi_m(z), 
\ee
where the potential is given by
\be
V(z) =\frac{15k^{2}}{8[g(z)]^{2}}\left[\theta
(z-z_{1})-\theta (z-z_{2})\right]  \nonumber \\
-\frac{3k}{4g(z)}\left[ \delta (z-z_{1})-
\delta (z-z_{2})\right],
\ee
and
\be g(z)=\left\lbrace \matrix{ {kz_1+1} &{\rm for}& 0\leq z\leq
z_{1} \cr {kz+1} &{\rm for}& z_{1}\leq z\leq z_{2} \cr {kz_2+1}
&{\rm for}& z\geq z_{2}} \right.
\ee
with $z_1=z(y_1)$ and $z_2=z(y_2)$.

Solution to the Schr\"{o}dinger-like equation in Eq.(6) is a combination of
plane waves in the Minkowski region and Bessel functions in the
AdS region:

\be
\Psi_m=\left\lbrace \matrix{ A\cos{(m z)} &{\rm for}& 0\leq
z\leq z_{1} \cr \sqrt{ k^{-1} g(z)}\left[B_1
N_{2}\left(\frac{m}{k}g(z)\right)+B_2
J_{2}\left(\frac{m}{k}g(z)\right)\right] &{\rm for}& z_{1}\leq
z\leq z_{2} \cr C_1\cos(mz)+C_{2}\sin(mz) &{\rm for}& z\geq z_{2}}
\right.
\ee
where $J$ and $N$ are Bessel functions. The constants $A$, $B_1$
and $B_2$ obey two relations obtained from junction conditions at
the positive tension brane at $y=y_1$.
Similarly,
the relations between $B_1$, $B_2$ and $C_1$, $C_2$ are
given by the junction conditions at $y=y_2$.
Considering the normalized condition $\int |\Psi^2_m(z)| dz=1$ and
making use of the asymptotic behaviour of $\Psi$ at $z\rightarrow
\infty$, we have,

\be
\pi\left(C_1^2+C_2^2\right)=1.
\ee

Therefore, $A$ can be formulated as
{\footnotesize
\begin{eqnarray}
&&{1\over \pi A^2}\left[N_2 \left({m\over
k}g_1\right)J_1\left({m\over k}g_1\right)
-N_1 \left({m\over k}g_1\right)J_2\left({m\over k}g_1\right)\right]
\left({g_1\over g_2}\right)\nonumber\\
&=&\left[J_1\left({m\over k}g_1\right) \cos{(mz_1)}+
J_2 \left({m\over k}g_1
\right)\sin{(mz_1)}\right]^2 \left[N_1^2\left({m\over k}g_2\right)+N_2^2
\left({m\over k}g_2\right)\right]\nonumber \\
&-&2\left[J_1 \left({m\over k}g_1\right)
\cos{(mz_1)}+J_2\left({m\over k}g_1
\right) \sin{(mz_1)}\right]\left[N_1\left({m\over k}g_1\right) \cos{(mz_1)}
+N_2 \left({m\over k}g_1\right)\sin{(mz_1)}\right]\nonumber\\
&&\left[N_1 \left({m\over k}g_2\right)J_1\left({m\over k}g_2\right)
+N_2\left({m\over k}g_2\right)J_2\left({m\over k}g_2\right)\right]\nonumber \\
&+&\left[N_1 \left({m\over k}g_1\right)\cos{(mz_1)}
+N_2\left({m\over k}g_1\right)\sin{(mz_1)}\right]^2\left[J_1^2
\left({m\over k}g_2\right)+J_2^2\left({m\over k}g_2\right)\right].
\end{eqnarray}}

Now we calculate the static gravitational
potential between two unit masses placed on the $z=z_1$ positive
tension brane at a distance $r$ from each other. This potential is
generated by the exchange of the massive modes. Following Ref.
\cite{GW1} (also see Ref. \cite{GT} for a more extensive
discussion), we have
\begin{eqnarray}
V(r)&=&-G_5\int_0^{\infty} dm {e^{-mr}\over r}\Psi_m^2(z=z_1)\nonumber \\
&=&-G_5\int_0^{\infty} dm {e^{-mr}\over r} A^2 \cos^2 {(mz_1)},
\end{eqnarray}
where $G_5$ is 5D Newton constant.

In Fig.2 and Fig.3 we plot
$A^2\cos^2 {(mz_1)}$ vs $m z_1$,
from which one can see that for $mz_1\gg 1$,
$A^2\cos^2{(mz_1)}$ is a periodic function and independent of the
value of $z_2/z_1$. Eq.(12) shows that the massive KK 
modes effectively modify Newton law on the positive tension brane
at various length scale.

We now present results in the case of $kz_1\gg 1$. 
Conveniently we rewrite the integral in (12) into three parts:
\begin{eqnarray}
V(r)&=&V_1(r)+V_2(r)+V_3(r)\nonumber\\
&=&-G_5\int_0^{z_1^{-1}} dm {e^{-mr}\over r}A^2 \cos^2(m z_1)
-G_5\int_{z_1^{-1}}^{z_2^{-1}} dm {e^{-mr}\over r}A^2 \cos^2(m z_1)
\nonumber \\
&-&G_5\int_{z_2^{-1}}^{\infty} dm {e^{-mr}\over r}A^2 \cos^2(m z_1).
\end{eqnarray}

The constant $A$ in the first term of right-handed side of Eq.(13) 
{\it i.e.} $V_1(r)$ can be evaluated by 
using the series expansion of Bessel functions with the arguments $mg_1/k$
and $mg_2/k$:

 \be
 A^2\cos^2{(mz_1)}\simeq {1\over
\pi}{49\over 64} {(z_2/z_1)^3\over 1+{9\over 4}(mz_1)^2
(z_2/z_1)^6}.
 \ee
 Substituting (14) into (13), we get

  \be
 V_1(r)\simeq -{2G_N\over \pi r} \int_0^{r_1/z_1} dx {1\over 1+x^2}
e^{-xr/r_1},
 \ee
 where $x={3\over 2}mz_1(z_2/z_1)^3\equiv mr_1$ and $G_N=G_5/2z_1$
is the renormalized 4D Newton constant.

We see that $V(r)=-G_N/r$ for $r\ll r_1$, i.e., the 4D Newton law
in this case, however $V(r)=-\frac{G_Nr_1}{\pi r^2}$ for $r\gg
r_1$ which has the form of Newton law of 5D gravity with a
renormalized gravitational constant. This result is similar to
that of Ref.[5], except for the different normalization of
$G_N$ and definition of $r_1$. This is because in the limit of
$z_1\rightarrow 0$ our model approaches to
the GRS model\cite{GRS}.

The second term in the right hand side of (13), {\it i.e.}$V_2(r)$ is 
similar 
to the contribution of the continuum modes to the gravitational potential
in the Randall-Sundrum model:

 \be
 A^2\cos^2(mz_1)\simeq mz_2.
 \ee
 It gives rise to a correction to Newton law at short distance,
 \be V_2(r)\simeq -{G_N\over
r}{z_1 z_2\over r^2}.
 \ee
  For $r\ll \sqrt{z_1 z_2}$, this term
will dominate the gravitational potential $V(r)$.

 Using the asymptotic expansion of Bessel functions, we obtain the constant 
$A$ in $V_3(r)$,
 \be
A^2\cos^2{(mz_1)}\simeq {1\over \pi}\cos^2{(mz_1)}.
 \ee
 Substituting (18) into (13), we have
\begin{eqnarray}
V_3(r)&\simeq &-{2G_N\over \pi r}\int_{z_1/z_2}^{\infty}dx\, \,
  e^{-xr/z_1}
\cos^2{x}\nonumber\\
&=&-{G_N\over r}{z_1\over \pi r}\left[1+{(r/z_1)\over 4+(r/z_1)^2}
\left({r\over z_1}\cos\left(2{z_1\over z_2}
\right)-\sin\left(2{z_1\over z_2}
\right)\right)\right]
e^{-r/z_2} .
\end {eqnarray}
One can see that $V(r)\simeq -{G_N z_1\over r^2}e^{-r/z_2}$ for
$r\ll z_1$. This is the Yukawa-type potential with a renormalized
gravitational constant, however for $r\gg z_1$, $V(r)\simeq -{G_N 
z_1\over
r^2}[1+\cos{(2z_1/z_2)}]$. Comparing (19) with (17)
shows that for all values of $z_1$ and $z_2$, $V_3 < V_2$. This is
 different from Ref. \cite{LMW} and this difference comes from
the different normalization condition. 
In our 
model the contribution of massive resonant modes is not the leading
contribution to the Newton gravitional potential at small scale.
In Fig.4 we show the comparison between the two models at small scale.

 The sum of $V_1(r)$, $V_2(r)$ and $V_3(r)$ gives rise to the
 gravitational potential on the brane located 
at $z=z_1$ 
\begin{eqnarray}
V(r)&=&-f(r){G_N\over r}\nonumber\\
&\simeq& -{2G_N\over \pi r} \int_0^{r_1/z_1} dx {1\over 1+x^2} e^{-xr/r_1}
\nonumber\\
&&-{G_N\over r}{z_1 z_2\over  r^2}\nonumber\\
&&-{G_N\over r}{z_1\over \pi r}\left[1+{(r/z_1)\over 4+(r/z_1)^2}
\left({r\over z_1}\cos\left(2{z_1\over
z_2}\right)-\sin\left(2{z_1\over z_2} \right)\right)\right]
e^{-r/z_2},
\end{eqnarray}
where $f(r)$-1 represents the deviation from the standard 4D Newton
gravitation. In Fig.5 we plot $f(r)$  as a function of $r$.
One can see that only in the region of
 $\sqrt{z_1
z_2}\ll r\ll r_1$
the gravitational potential for two unit masses
on the positive tension brane is described by the 4D Newton law
 \be V(r)
 \simeq -{G_N\over r}.
 \ee
 In the case of $r\ll \sqrt{z_1 z_2}$, the
term $z_1 z_2/r^2$ will be dominative. On the other hand for
 $r\gg r_1$, we have the 5D Newton gravitation with a
renormalized Newton gravitational constant which is smaller than that in 4D.

In comparison with the observational data, we take
$r_1\sim 10^{28}$cm, i.e.,
the present horizon size of the Universe.
Defining $\sqrt{z_1 z_2}= l$ in the unit of centimeter, in Fig.6 we plot
$ky_1$ and $ky_2$ as function of $l$.
The updated observation \cite{HS} put a limit 
$l\sim 0.2$mm, from which and Fig.6 we obtain that 
$ky_1\simeq 60$ and $ky_2\simeq 90$. As shown in Ref.[8], to have these 
values it does not require a strong fine tuning of the model parameters. 

In summary, we have proposed a brane setup where extra dimension has a
non-factorized geometry and in the fifth dimension Minkowski regions 
and the AdS regions are separated by branes. In our 
model there are both the graviton-like resonant state which generates the
``quasi-localized gravity" and the resonant states of the massive
KK mode. The latter one, however in comparison with the continuum modes, 
is
not the leading contribution to Newton gravitational potential at
small scale, and also differs from that in Ref. \cite{LMW}. Our model
in this paper
provides a picture where gravitational interactions
at different length scales possess different behaviors. In
particular Newton coupling constant depends on the parameters of the
configuration. Finally, we have shown that our model is in consistent 
with the 
updated observation without a strong fine tuning of the model parameters.

\textbf{Acknowledgments}

 This work is supported in part by National Natural Science Foundation of
China under Grant Nos. 10047004 and 19835040, and also supported
by Ministry of Science and Technology of China under Grant No. NKBRSF
G19990754.

\vspace*{10mm}

\begin{figure}[ht]

\begin{center}

\setlength{\unitlength}{0.1in}

\begin{picture}(30,12)

\put(-8,5){\line(1,0){6}}

\put(0,5){\line(1,0){8}}

\put(10,5){\line(1,0){10}}

\put(22,5){\line(1,0){8}}

\put(32,5){\line(1,0){5}}

\put(15,5){\line(0,1){0.5}}

\put(-1,5){\circle * {2}}

\put(9,5){\circle{2}}

\put(21,5){\circle{2}}

\put(31,5){\circle * {2}}

\put(-1.5,2){\makebox(1,1)[c]{$-$}}

\put(8.5,2){\makebox(1,1)[c]{$+$}}

\put(20.5,2){\makebox(1,1)[c]{$+$}}

\put(30.5,2){\makebox(1,1)[c]{$-$}}

\put(-2,8){\makebox(1,2)[c]{$-y_2$}}

\put(8,8){\makebox(1,2)[c]{$-y_1$}}

\put(20.5,8){\makebox(1,1)[c]{$y_1$}}

\put(30.5,8){\makebox(1,1)[c]{$y_2$}}

\put(14,2){\makebox(1,1)[c]{$y=0$}}

\put(-7,6){\makebox(1,3)[c]{$Min$}}

\put(4,6){\makebox(1,3)[c]{$AdS$}}

\put(14,6){\makebox(1,3)[c]{$Min$}}

\put(25,6){\makebox(1,3)[c]{$AdS$}}

\put(35,6){\makebox(1,3)[c]{$Min$}}

\end{picture}

\caption {Illustration of our 5D model consisted of four parallel 
$3$-branes with
Minkowski and AdS spacetime alternated and a 
$z_2$ symmetry in the fifth dimension $y$. }

\end{center}

\end{figure}
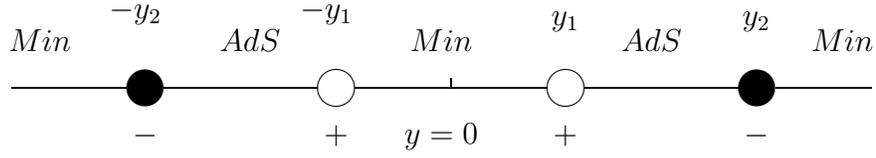

\begin{figure}

\begin{center}

\mbox{\epsfig{file=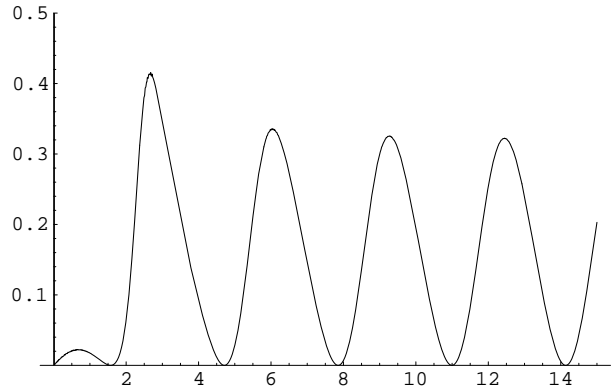,width=8cm}}

\caption { Plot of $A^2\cos^2 {(mz_1)}$ (y-axis){\it vs}
$mz_1$ (x-axis) for $z_2/z_1=100$. }

\end{center}

\end{figure}

\begin{figure}[ht]

\begin{center}

\mbox{\epsfig{file=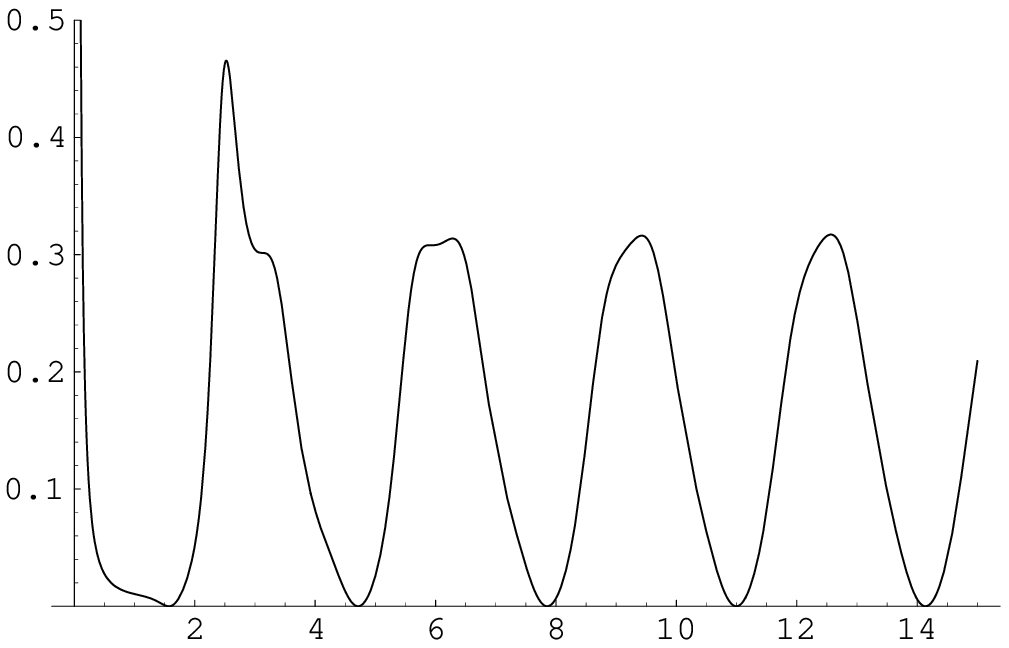,width=6cm}\epsfig{file=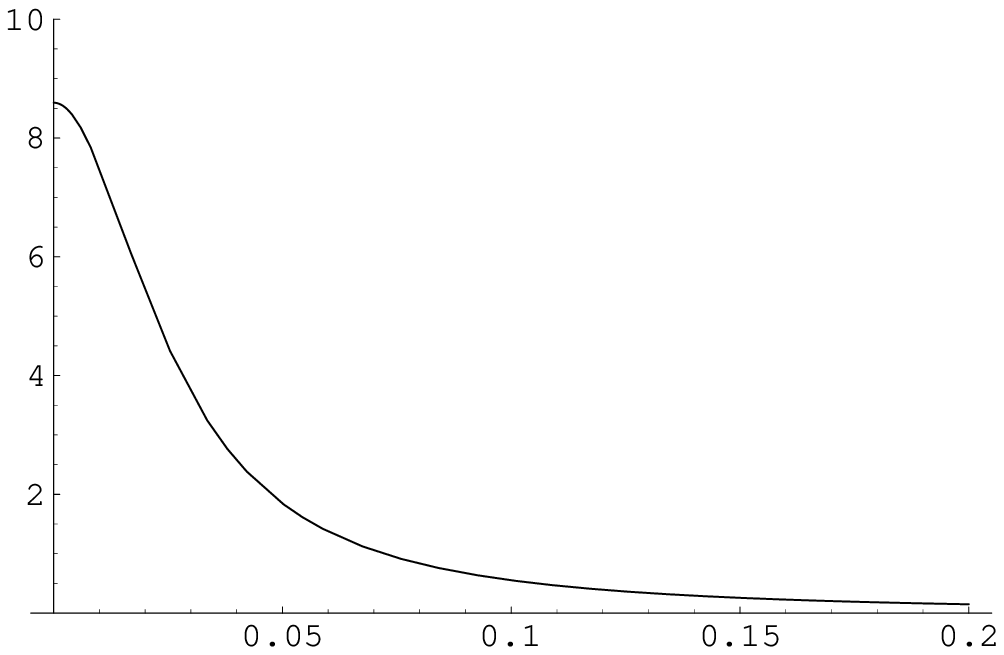,width=6cm}}

\caption { Plot of $A^2\cos^2 {(mz_1)}$ (y-axis) {\it vs}
$mz_1$ (x-axis) for $z_2/z_1=3$. (a) and (b) are, respectively,
for $mz_1 = [0, 15]$ and  $mz_1 = [0, 0.2]$. }

\end{center}

\end{figure}

\begin{figure}[ht]

\begin{center}

\mbox{\epsfig{file=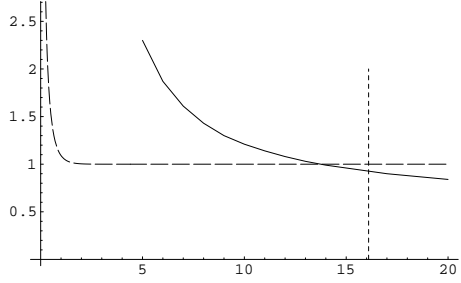,width=6cm}}

\caption {Comparison of the $f(r)$ function in our model
(the solid line) with that (the long-dash-line) of Ref. \cite {LMW}
where $z_c$ corresponding to $z_1$ here is taken to be $10^{-9}$cm. The 
x-axis is $\ln({r\over z_1})$ and ${z_2\over z_1}\sim 10^{14}$ is
assumed. }

\end{center}

\end{figure}

\begin{figure}[ht]

\begin{center}

\mbox{\epsfig{file=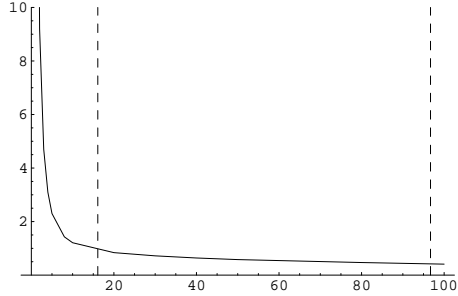,width=6cm}}

\caption {The solid line is $f(r)$ and between the two dash lines
 is region where the 4D Newton gravitation is approximately valid.
The x-axis is $\ln({r\over z_1})$( with ${z_2\over z_1}\sim 10^{14}$).
}

\end{center}

\end{figure}

\begin{figure}[ht]

\begin{center}

\mbox{\epsfig{file=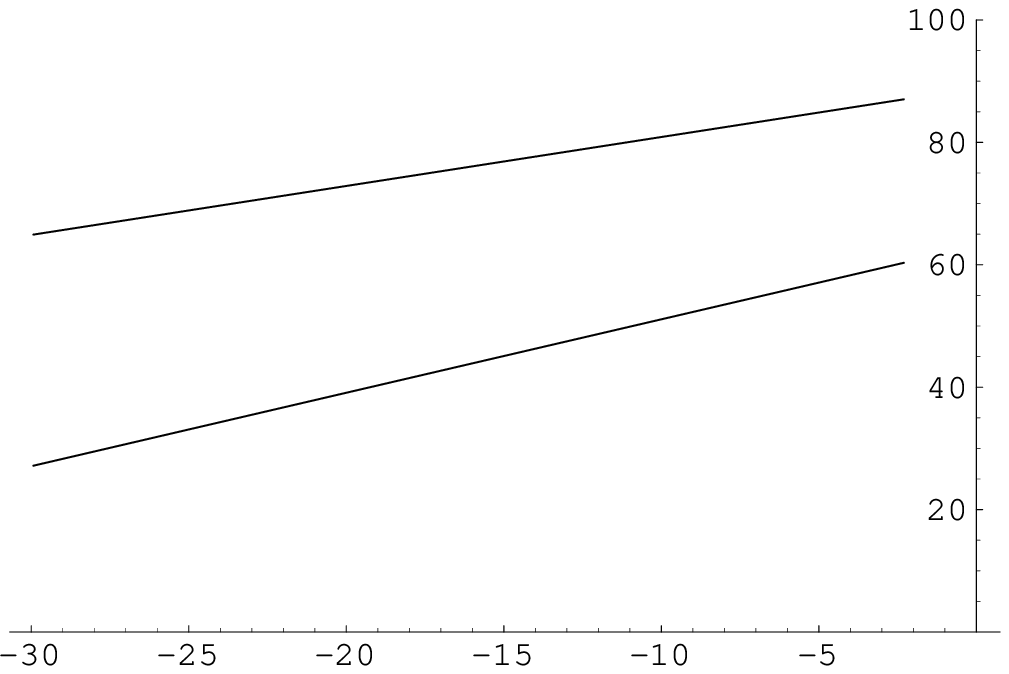,width=6cm}}

\caption {Plot of $ky_2$ (the upper curve) and
 $ky_1$ (the lower curve) {\it vs} $l$ in the region of $10^{-12}\sim 
10^{-1}$cm. The
x-axis is $\ln l$.
}

\end{center}

\end{figure}


\begin{thebibliography}{99}

\bibitem{HW} P. Horava and E. Witten, Nucl. Phys. \textbf{B460} 506 (1996);

\textbf{B475} 94 (1996).

\bibitem{WL}E. Witten, Nucl. Phys. \textbf{B471} 135 (1996);\\
J.D. Lykken, Phys. Rev. \textbf{D54} 3693 (1996)

\bibitem{ADD} N. Arkani-Hamed, S. Dimopoulos and G. Dvali, Phys. lett.

\textbf{B249} (1998) 263.

\bibitem{RS} L. Randall and R. Sundrum, Phys. Rev. Lett. \textbf{83} (1999)
3370.

\bibitem{GRS} R. Gregory, V.A. Rubakov and S.M. Sibiryakov, hep-th/0002072,
Phys. Rev. Lett. \textbf{84} (2000) 5928.

\bibitem{KR} I.I. Kogan and G.G. Ross, hep-th/0003074, Phys. Lett. \textbf{B485}
 (2000) 255.

\bibitem{K} N. Kaloper, Phys. lett. \textbf{B474} (2000) 269.\\
I.I. Kogan, S. Mouslopoulos, A. Papazoglou and G.G Ross,
hep-th/0006030, Nucl. Phys. \textbf{B595} (2001) 225.

\bibitem{YS} Y.S. Piao, X.M. Zhang and Y.Z. Zhang, hep-th/0104020,
 Phys. Lett. \textbf{B512} (2001) 1.

\bibitem{ADDK} N. Arkani-Hames, S. Dimopoulos, G. Dvali and N. Kaloper,
hep-ph/9911386, JHEP 12 (2000) 010.



\bibitem{CEH} C. Csaki, J. Erlich and T.J. Hollowood, Phys. Rev. Lett
\textbf{84} (2000) 5932.\\
G. Dvali, G. Gabadadze and M. Porrati, Phys. lett.
\textbf{B484} (2000) 112.



\bibitem{LMW} J. Lykken, R.C. Myers and J. Wang, hep-th/0006191, JHEP 9 (2000)
009.



\bibitem{GW1} L. Randall and R. Sundrum, Phys. Rev. Lett. \textbf{83}
(1999) 4690.



\bibitem{GT} J. Garriga and T. Tanaka, Phys. Rev. Lett. \textbf{84} (2000)
2778.\\
S.B. Giddings, E. Katz and L. Randall, JHEP \textbf{3} (2000) 023.



\bibitem{HS} C.D. Hoyle, U. Schmidt, B.R. Heckel, E.G. Adelberger, 
J.H. Gundlach,
D.J. Kapner and H.E. Swanson, Phys. Rev. Lett. \textbf{86} (2001) 1418



\end{thebibliography}
\end{document}